\documentclass[12pt]{article}
\usepackage{jheppub}
\usepackage{graphicx,color}
\usepackage[dvipsnames]{xcolor}
\usepackage{braket}
\usepackage{soul}
\usepackage{amssymb,amsmath}
\usepackage{bbm}
\usepackage{array}
\usepackage{booktabs}
\usepackage{multirow}

\usepackage{dsfont}
\allowdisplaybreaks
\usepackage[]{hyperref}
\hypersetup{
    colorlinks = true,
    allcolors = {MidnightBlue}
}
\allowdisplaybreaks

\def\id{{\mathbbm{1}}}

\def\d{\mathrm{d}}
\def\i{\mathrm{i}}

\def\id{\mathbbm{1}}

\title{The many facets of a hyperbolic tetrahedron: open and closed triangulations of 3d gravity}

\author{Daniel L. Jafferis}
\author{and Diandian Wang}

\affiliation{Jefferson Physical Laboratory, Harvard University, Cambridge, MA 02138, USA}
\emailAdd{jafferis@g.harvard.edu}
\emailAdd{diandianwang@fas.harvard.edu}

\date{\today}

\begin{document}

\abstract{
    We study a model of 3d gravity relevant to the open sector of a CFT ensemble. The quantum theory is the open Virasoro TQFT, obtained by restricting the full open-closed Virasoro TQFT to a subclass of admissible manifolds. We show that it computes gravitational path integrals on compact regions with fixed-length boundary conditions for states above the black hole threshold, and fixed-angle boundary conditions for states below the threshold. Focusing on a special class of manifolds involving only boundary Wilson loops, we further show that the relation between Conformal Turaev-Viro theory and the diagonal sector of two copies of Virasoro TQFT arises naturally from an open-closed duality. 
}

\maketitle

\section{Introduction}
Pure gravity in three dimensions (with a negative cosmological constant) occupies a special place in the study of quantum gravity and holography. Although it lacks local propagating graviton degrees of freedom, it retains a rich spectrum of global and topological phenomena \cite{Achucarro:1986uwr,Witten:1988hc}. Recently, it has become clear that 3d gravity is closely related to an ensemble of 2d CFTs, both in terms of OPE coefficients \cite{Kraus:2016nwo,Cardy:2017qhl,Collier:2019weq,Belin:2020hea,Belin:2021ryy,Anous:2021caj,Chandra:2022bqq,Collier:2023fwi,Belin:2023efa,deBoer:2023vsm,Collier:2024mgv,deBoer:2024mqg,Jafferis:2024jkb,Chandra:2025fef,Belin:2026pko,Wang:2025jgo} and spectral data \cite{Maloney:2007ud,Keller:2014xba,Cotler:2020ugk,Maxfield:2020ale,DiUbaldo:2023qli,Boruch:2025ilr}, and in turn, aspects of the CFT help illuminate 3d gravity. In particular, via a non-rational analog of the Chern-Simons/Wess-Zumino-Witten correspondence \cite{Witten:1988hf}, Virasoro conformal blocks play a key role in establishing 3d gravity on hyperbolic manifolds as a topological quantum field theory, namely Virasoro TQFT \cite{Collier:2023fwi,Collier:2024mgv}. 

A natural next step is to extend this relation from ordinary CFTs to boundary conformal field theories (BCFTs). On the gravitational side, this requires enlarging the class of allowed bulk configurations to include end-of-the-world (EOW) branes, which provide the holographic realization of conformal boundaries \cite{Takayanagi:2011zk,Fujita:2011fp,Wang:2025bcx}. This model correctly reproduces universal dynamics of BCFTs \cite{Kusuki:2021gpt,Numasawa:2022cni}, including the $g$-function dependence, which arises from of a topological nature of the EOW brane action \cite{Geng22,Wang:2025bcx}. At the quantum level, one obtains an open-closed extension \cite{Lazaroiu:2000rk,Lauda:2005wn,Moore:2006dw} of Virasoro TQFT on hyperbolic manifolds \cite{Jafferis:2025yxt}, and a matrix model description on certain off-shell manifolds \cite{Collier:2023cyw,Jafferis:2025jle}. The open sector is related to a chiral copy of Virasoro TQFT via the doubling trick \cite{Hung:2025vgs}, but the global aspects differ, such as the 2d topological action carried by the EOW brane and the difference in the mapping class group \cite{Jafferis:2025jle}. 

It is nevertheless interesting to isolate the purely open sector within this broader framework. From the BCFT point of view, this sector captures observables built entirely from boundary degrees of freedom, while from the 3d point of view it probes a distinguished class of geometries supported by sufficient EOW branes. As we will demonstrate in this work, this purely open theory is already nontrivial and exhibits clean features that are otherwise obscured by structures like modular transformations in the full open-closed system.

As recently shown by Hartman \cite{Hartman:2025ula}, with an appropriate normalization for trivalent junctions of Virasoro Wilson graphs, at large $c$, the diagonal sector of two copies of Virasoro TQFT reproduces the classical gravitational action in a compact region of spacetime. This generalizes the analogous relation known in the discrete-spectrum case \cite{Barrett:2004im}. The boundary of the compact region consists of pleated surfaces, with boundary conditions fixing the lengths of a number of geodesic circles, each related to the weight of a Wilson line that is above the black hole threshold. We find an analog of this in the open sector, where the open Virasoro TQFT computes the gravitational action of a compact region whose boundary consists of both EOW branes and pleated surfaces. Moreover, the open Virasoro TQFT computes a gravitational path integral with fixed-length boundary conditions for above-threshold states and fixed-angle boundary conditions for below-threshold states.

At the quantum level, the boundary Wilson lines in the open TQFT can form closed loops. Integrating over the weights of a closed Wilson loop against the Cardy density turns it into (the open version of) the $\Omega$ loop, which has a nice geometric interpretation \cite{Burnell:2010mx}. The procedure that turns the manifold containing the open $\Omega$ loop into one that does not is called the annular surgery \cite{Jafferis:2025yxt}. We will review how gluing $6j$ manifolds followed by performing annular surgeries is equivalent to a dual picture in which the same manifolds are built by tetrahedral decomposition.

We then focus on a special class of manifolds constructed within this purely open TQFT, involving only Wilson loops on EOW branes (no trivalent junctions), and use an open-closed duality to provide a natural explanation for the relation between Conformal Turaev-Viro (CTV) theory \cite{Turaev:1992hq,Hartman:2025cyj} and two copies of Virasoro TQFT with scalar Wilson graphs:
\begin{align}\label{eq:CTVVir0}
Z_{\mathrm{CTV}}[\tilde{M}_E, \tilde{\Gamma}(\mathbf{P})]=\prod_{i}\left(\int_0^{\infty} \d P_i \,\mathbb{S}_{P_i P_i^{\prime}}[\id]\right)\left|\hat{Z}_{\mathrm{Vir}}[\tilde{M}_E, \tilde{\Gamma}(\mathbf{P}^{\prime})]\right|^2.
\end{align}
To do this, we first establish an equivalence between the open Virasoro TQFT partition function of the special class of manifolds and the CTV partition function; we then use the open-closed duality, which involves a Fourier transform, to relate it to the scalar sector of the closed Virasoro TQFT, namely two copies of Virasoro TQFT with diagonal weights. The notational details of this equation will be explained in the main text.

When the weights of these Wilson loops are below the threshold, the open Virasoro TQFT partition function computes a fixed-angle path integral, and, consequently, so does the corresponding CTV partition function. This provides an alternative perspective on the distinction between fixed-length and fixed-angle boundary conditions for computing OPE statistics in the closed sector.

The plan for the rest of the paper is as follows. In Section~\ref{sec:open}, we study the purely open model, both from the perspective of BCFT partition functions, where the bulk manifolds have asymptotic boundaries, and from the perspective of OPE statistics, where the bulk manifolds have finite (unrenormalized) volumes. In Section~\ref{sec:trig}, we explain how the bulk manifolds with finite boundaries can be built using triangulation and how the open Virasoro TQFT computation is equivalent to tetrahedral decomposition. The difference between above-threshold and below-threshold states is also clarified there. In Section~\ref{sec:oc}, we review and apply an open-closed duality to relate certain partition functions of the purely open model to certain partition functions in the closed sector, reproducing a relation between CTV and two copies of Virasoro TQFT. Finally, we conclude in Section~\ref{sec:disc} with a summary and discussion.

\section{Purely open ensemble}\label{sec:open}

The essence of the AdS/BCFT correspondence \cite{Takayanagi:2011zk,Fujita:2011fp} is to relate gravity with EOW branes to BCFT partition functions. For 2d BCFTs, it states that
\begin{align}\label{eq:dictionaryasymp}
    \overline{Z[\Sigma_{g,n}]} = \sum_{M}Z_{\rm grav}[M], \quad \partial M=Q\cup \Sigma_{g,n},    
\end{align}
where $Z[\Sigma_{g,n}]$ is the BCFT partition function on the genus-$g$ Riemann surface with $n$ borders, and the sum is over all manifolds whose boundary $\partial M$ is the union of the asymptotic boundary $\Sigma_{g,n}$ and some other (possibly disconnected) 2d surface $Q$, which we call the EOW brane. We have included a bar here, which denotes statistical averaging, because the existence of wormholes in gravity suggests an ensemble of theories rather than a single one \cite{Maldacena:2004rf,Chandra:2022bqq}.

For the purpose of this work, we will only consider bulk manifolds that admit saddles under the action specified below, and we refer to such manifolds as hyperbolic manifolds. This definition reduces to the usual definition of hyperbolicity when $\partial M=\Sigma_{g,0}$, i.e., in the absence of EOW branes. 

As explained in \cite{Belin:2026pko}, it is sufficient to consider a single connected partition function, because products of partition functions
\begin{align}\label{eq:Zprod}
    \overline{Z[\Sigma_{g_1,n_1}]Z[\Sigma_{g_2,n_2}]\dots}
\end{align}
can be obtained from $\overline{Z[\Sigma_{g,n}]}$ for some $g$ and $n$ by taking a limit in the moduli space (as long as we consider only hyperbolic manifolds). The limit is taken so that some geodesic circles and/or intervals pinch off. In this limit, the only hyperbolic bulk manifolds contributing to $\overline{Z[\Sigma_{g,n}]}$ are handlebody-like at the pinching circles and intervals, by which we mean manifolds where the pinching circles are contractible in the bulk and the pinching intervals are contractible in the bulk with the restriction that the ends of the interval stay on the EOW brane, by a version of the Schlenker-Witten theorem \cite{Schlenker:2022dyo,Wang:2025bcx}. (Equivalently, a pinching circle bounds a disk, and a pinching interval together with an interval on the EOW brane bounds a disk.) In the pinching limit of the Riemann surface, the corresponding handles themselves degenerate, and the resulting manifolds contribute to \eqref{eq:Zprod}, upon multiplication by appropriate factors due to the conformal anomaly  \cite{Cardy:2017qhl}. Importantly, the bulk manifolds can still be connected, and if so, they are connected contributions to \eqref{eq:Zprod} and often referred to as Euclidean wormholes.

In this section, we consider the simple model with only open states, as advocated in the introduction. We will first specify the bulk theory by stating the action and boundary conditions in Section~\ref{ssec:actionasymp}. We then discuss how the partition function is expressed in terms of purely open data in Section~\ref{ssec:channel}. Next, we explain why it is sufficient to focus on the OPE statistics, and write down the dictionary for computing the OPE statistics in the fixed momentum basis in Section~\ref{ssec:actfinite}. Finally, in Section~\ref{ssec:openV}, we describe an equivalent way to formulate the problem in terms of Wilson graphs of the open Virasoro TQFT.

\subsection{Action: asymptotic boundary}\label{ssec:actionasymp}

The dictionary \eqref{eq:dictionaryasymp} relates path integrals on asymptotically AdS manifolds to BCFT partition functions on Riemann surfaces with fixed moduli. We now review the action of the gravity model that has been assumed in this relation. 

For manifolds with asymptotic boundary, the Euclidean action of the gravity theory is given by \cite{Takayanagi:2011zk,Fujita:2011fp}
\begin{align}\label{eq:action_tot}
    I =
    I_{\rm bulk} +  I_{\rm brane}+I_{\rm asymp},
\end{align}
where
\begin{align}
    I_{\rm bulk}=-\frac{1}{16\pi G_{N}} \int_{M} \sqrt{g} \,(R +2)
\end{align}
is the Einstein-Hilbert action for the bulk of the manifold $M$, with the cosmological constant set to $\Lambda=-1$, 
\begin{align}\label{eq:action_EOW}
    I_{\rm brane}=- \frac{1}{8\pi G_{N}}\sum_a \int_{Q_a} \sqrt{h}\, (K-T_a)
\end{align}
is the action for the EOW brane $Q=\sqcup_a Q_a$ with $a$ labeling its different connected components \cite{Karch:2000ct,Karch:2000gx}, $K$ is the trace of the extrinsic curvature $K=\nabla_\mu n^\mu$ with the normal $n^\mu$ pointing outwards, $h$ is the determinant of the induced metric, and $T_a$ is the tension of the connected brane component $Q_a$, and 
\begin{align}
I_{\rm asymp}=-\frac{1}{8 \pi G_{N}} \int_{A} \sqrt{h}\,(K-1)
\end{align}
is the action at the asymptotic boundary $A$, which includes both the Gibbons-Hawking-York term and the counterterm.

For each connected component of the brane $Q_a$, the tension $T_a$ is related to the $g$-function associated to the boundary condition of the BCFT at $\partial Q_a\subseteq \partial A$ via \cite{Takayanagi:2011zk,Fujita:2011fp}
\begin{align}\label{eq:gandT}
    \log g_a = \frac{1}{4G_N}{\rm arctanh} \,T_a= \frac{c}{6}{\rm arctanh} \, T_a. 
\end{align}
For on-shell manifolds, including a topological term is equivalent to including a tension term \cite{Geng22,Wang:2025bcx}, so it is equivalent to the following action for the EOW brane:
\begin{align}
    I_{\rm brane}=- \sum_a\chi_a \log g_a- \frac{1}{8\pi G_N} \int_{Q} \sqrt{h}\, K.
\end{align}
The first term is topological, also known as the Marolf-Maxfield action \cite{Marolf:2020xie}, and the second term is the same as \eqref{eq:action_EOW} but with the tension set to zero. In practice, it is much simpler to work with EOW branes with zero tension and use the topological term to keep track of the dependence of the action on $g_a$ and therefore $T_a$, so we will do so throughout the rest of this work.

At the asymptotic boundary $A$, we impose the usual Dirichlet boundary conditions. In particular, we fix the moduli of the bordered Riemann surface $\Sigma_{g,n}$. At the EOW branes, we impose Neumann boundary conditions, requiring that the extrinsic curvature tensor vanish.

\subsection{Channel decomposition}\label{ssec:channel}

Consider the BCFT partition function on a bordered Riemann surface $\Sigma_{g,n}$ with at least one border ($n\ge1$) so that open strings can end. By definition, the disk partition function $Z[\Sigma_{0,1}]$ is the $g$-function.  On $\Sigma_{0,2}$, the annulus partition function provides the spectral density. In the interest of studying nontrivial OPE statistics, it therefore suffices to consider channel decompositions for $2g+n\ge3$.

Given $Z[\Sigma_{g,n}; \Omega]$, where $\Omega$ denotes the moduli of the Riemann surface, we pick a purely open channel and expand it in the $P$ basis as
\begin{align}\label{eq:Zexpand}
\overline{Z[\Sigma_{g,n}; \Omega]}=\int \mathrm{d}^{6g+3n-6} P  \overline{
\prod_{e^{(ab)}_I \in \Gamma_{\mathcal{C}}} 
(g_{a}g_b)^{-1/2}\rho^{(ab)}(P_I) \prod_{v^{(abc)}_{IJK} \in \Gamma_{\mathcal{C}}} B^{(abc)}_{IJK}
} 
\,
\mathcal{F}_{g,n}^{\mathcal{C}}(\mathbf{P} ; \Omega),
\end{align}
where $\mathcal{F}_{g,n}^{\mathcal{C}}(\mathbf{P} ; \Omega)$ is the BCFT Virasoro conformal block, to be defined with more precision shortly. Each $e_I^{(ab)}$ labels an edge of the graph, and each $v_{IJK}^{(abc)}$ labels a vertex.

Here, $\rho^{(ab)}$ is the spectral density of boundary operators in the open Hilbert space $\mathcal{H}^{(ab)}_{\rm open}$ defined on intervals extending between boundaries labeled by $a$ and $b$, and $B_{IJK}^{(abc)}$ is the three point function of boundary operators on the disk, or the boundary-to-boundary OPE coefficient. The convention for $B$ is such that
\begin{align}\label{eq:datapics}
   B_{IJK}^{(abc)}~ = ~\vcenter{\hbox{\includegraphics[height=2cm]{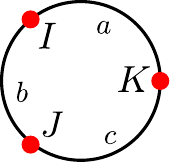}}}~,
\end{align}
where the red dots are boundary operator insertions, and the intervals between them are labeled with generally different boundary conditions. Via (the open version of) the operator-state correspondence, it is equivalent to
\begin{align}
    B_{IJK}^{(abc)}~=~\vcenter{\hbox{\includegraphics[height=2cm]{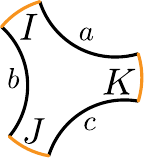}}}~=~\text{``open pair of pants''},
\end{align}
where the orange intervals are where the open states reside, and the black intervals are physical boundaries with chosen boundary conditions $a,b,c$.

The appearance of the factor of $(g_ag_b)^{-1/2}$ for each edge of the graph comes from the choice for the normalization of the two-point function (using the convention as \cite{Numasawa:2022cni,Wang:2025bcx,Hung:2025vgs}), which is fixed by
\begin{align}\label{eq:Bidnorm}
    B^{(aba)}_{IJ\id}=\sqrt{g_{a}g_b}\, \delta_{IJ}.
\end{align}
In other words, each contraction of a pair of lower indices ($I,J$) appearing in the product of OPE coefficients is carried out by the inverse metric $(g_ag_b)^{-1/2}\delta^{IJ}$.

To be more precise about the BCFT conformal block appearing in \eqref{eq:Zexpand}, we apply Cardy's doubling trick \cite{Cardy:1984bb}, 
\begin{align}
\Sigma_{g,n} \to \widehat{\Sigma}_{2g+n-1}, \quad \mathcal{C}\to\widehat{\mathcal{C}},\quad \Omega\to\widehat{\Omega},
\end{align}
which maps the bordered Riemann surface to a compact Riemann surface with genus $\widehat{g}=2g+n-1$ and moduli $\widehat{\Omega}$. The open channel $\mathcal{C}$ is mapped to closed channel $\widehat{\mathcal{C}}$, which is a decomposition of $\widehat{\Sigma}$ into ordinary pairs of pants rather than open pairs of pants (each open pair of pants joins with its mirror image to form an ordinary pair of pants). Then
\begin{align}
    \mathcal{F}_{g,n}^{\mathcal{C}}(\mathbf{P} ; \Omega)\equiv\mathcal{F}_{\widehat{g}}^{\widehat{\mathcal{C}}}(\mathbf{P} ; \widehat{\Omega}),
\end{align}
where the RHS is the Virasoro conformal block for the compact Riemann surface with genus $\widehat{g}$ in the channel $\widehat{\mathcal{C}}$. Note that the values of $\mathbf{P}$ are the same on both sides. The conformal blocks are delta-function normalizable with respect to the Verlinde inner product \cite{Verlinde:1989ua}. Using the normalization of \cite{Collier:2023fwi} for $\mathcal{F}_{\widehat{g}}^{\widehat{\mathcal{C}}}(\mathbf{P} ; \widehat{\Omega})$, we have
\begin{align}
\langle\mathcal{F}_{g,n}^{\mathcal{C}}(\mathbf{P}_1) |\mathcal{F}_{g,n}^{\mathcal{C}}(\mathbf{P}_2)\rangle=\frac{\delta^{(6g+3n-6)}(\mathbf{P}_1-\mathbf{P}_2)}{\rho_{g, n}^{\mathcal{C}}(\mathbf{P}_1)},
\end{align}
where
\begin{align}\label{eq:rhoC}
    \rho_{g,n}^{\mathcal{C}}(\mathbf{P})=\prod_{e^{(ab)}_I \in \Gamma_{\mathcal{C}}}
\rho_0(I)\prod_{v^{(abc)}_{IJK} \in \Gamma_{\mathcal{C}}} C_0({IJK}).
\end{align}
The two functions appearing here are the Cardy density and the universal OPE function \cite{Collier:2019weq} which is related to the DOZZ formula \cite{Dorn:1994xn,Zamolodchikov:1995aa}:
\begin{align}\label{eq:rho0C0def}
    \rho_0(P)&=4\sqrt{2}\sinh(2 \pi b P) \sinh ({2 \pi}{b^{-1}} P),
\\
    C_0(P_1, P_2, P_3)&=\frac{\Gamma_b(2 Q) \Gamma_b(\frac{Q}{2} \pm \i P_1 \pm \i P_2 \pm \i P_3)}{\sqrt{2}\, \Gamma_b(Q)^3 \prod_{k=1}^3 \Gamma_b(Q \pm 2 \i P_k)},
\end{align}
where $\Gamma_b(Q)$ is the double Gamma function, and the expression should be read as taking the product over all choices of the $\pm$ signs. To simplify notation, we have replaced momenta such as $P_I$ by just $I$, and will do so for other similar functions that appear in this work.

As an example, consider the bordered Riemann surface $\Sigma_{0,4}$ and decompose it into four open pairs of pants:
\begin{align}\label{eq:Sigma04example}
    \vcenter{\hbox{\includegraphics[height=3.5cm]{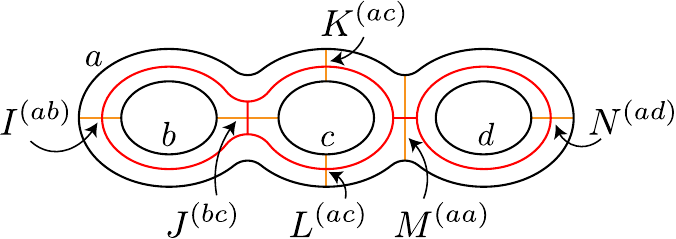}}}~.
\end{align}
The channel is specified by the trivalent graph (red). As stated earlier, each edge is associated to a spectral density $\rho^{(ab)}$, and each vertex is associated to an OPE coefficient $B_{IJK}^{(abc)}$. Equivalently, we can focus on the orange intervals that separate $\Sigma_{0,4}$ into open pairs of pants: we assign each interval a spectral density and each open pair of pants an OPE coefficient. In this example,
\begin{align}
    &\prod_{e^{(ab)}_I \in \Gamma_{\mathcal{C}}}(g_ag_b)^{-1/2} \rho^{(ab)}(P_I)
    \nonumber\\
    =
    &\,g_a^{-3}g_b^{-1}g_c^{-3/2}g_d^{-1/2}
     \rho^{(ab)}(P_I)
    \rho^{(bc)}(P_J)
    \rho^{(ac)}(P_K)
    \rho^{(ac)}(P_L)
    \rho^{(aa)}(P_M)
    \rho^{(ad)}(P_N),
\end{align}
and
\begin{align}
    \prod_{v^{(abc)}_{IJK} \in \Gamma_{\mathcal{C}}} B^{(abc)}_{IJK}
    =
    B^{(abc)}_{IJK}
    B^{(bac)}_{ILJ}
    B^{(aca)}_{KLM}
    B^{(aad)}_{MNN}.
\end{align}

It is worth emphasizing that the set of admissible contributing manifolds depends on the choice of boundary conditions. For $a\ne b$, the identity state $\mathbbm{1}$ is not in the open Hilbert space. In particular, this means that the open handlebody manifolds are forbidden unless all boundary conditions are identical. (Recall that by an open handlebody, we mean a manifold where all independent intervals at the asymptotic boundary are contractible in the bulk.)

In general, decomposition in a purely open channel is always possible for $2g+n\ge3$ (which is what we are considering). The problem of finding such a channel for $\Sigma_{g,n}$ is equivalent to finding a triangulation of $\Sigma_g$ with $n$ vertices. The Euler formula for the triangulated surface is given by
\begin{align}
    V-E+F=2-2g,
\end{align}
where $V$, $E$, and $F$ denote the numbers of vertices, edges, and faces, respectively. A triangulation has $3F=2E$, so we will get
\begin{align}
    F=2(2g+V-2),\quad E=3(2g+V-2).
\end{align}
This means that $\Sigma_{g,n}$ is formed by $2(2g+n-2)$ open pairs of pants, separated by $3(2g+n-2)$ open state cuts. 

As a second example, consider the decomposition of $\Sigma_{2,1}$ into open pairs of pants illustrated in the following diagram:
\begin{align}
   \vcenter{\hbox{\includegraphics[height=4cm]{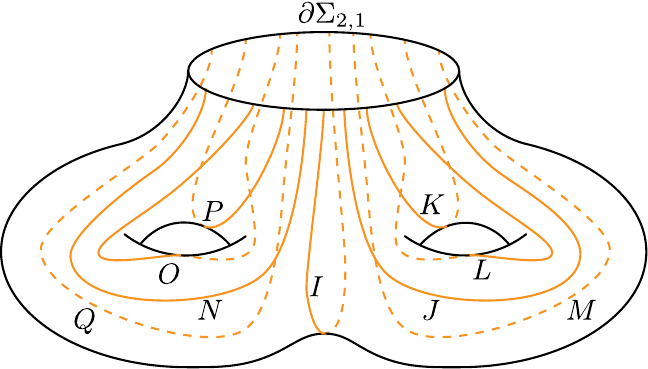}}}~.
\end{align}
The circle at the top of the diagram is the border of the Riemann surface, and the orange lines are open state cuts that separate the surface into open pairs of pants. Since there is just one border, we drop the boundary condition label for this example. Then
\begin{align}
    \prod_{e_I \in \Gamma_{\mathcal{C}}}g^{-1} \rho(P_I)
    =
    g^{-9}\rho(P_I)\rho(P_J)\rho(P_K)\rho(P_L)\rho(P_M)\rho(P_N)\rho(P_O)\rho(P_P)\rho(P_Q),
\end{align}
and
\begin{align}
    \prod_{v_{IJK} \in \Gamma_{\mathcal{C}}} B_{IJK}
    =
    B_{IMJ}B_{JLK}B_{KML}B_{INQ}B_{QPO}B_{ONP}
    .
\end{align}

The nontrivial part of the ensemble-averaged BCFT partition function \eqref{eq:Zexpand} lies in the statistical moments, as the conformal blocks are fixed by symmetry and have no statistical fluctuation. Furthermore, as the fluctuation in the spectrum is associated to non-hyperbolic manifolds \cite{Cotler:2020ugk,DiUbaldo:2023qli,Boruch:2025ilr,Jafferis:2025jle}, the spectral density is also fixed, i.e., 
\begin{align}\label{eq:pullout}
\overline{
\prod_{e^{(ab)}_I \in \Gamma_{\mathcal{C}}} 
(g_ag_b)^{-1/2}\rho^{(ab)}(P_I) \prod_{v^{(abc)}_{IJK} \in \Gamma_{\mathcal{C}}} B^{(abc)}_{IJK}
} 
=
\prod_{e^{(ab)}_I \in \Gamma_{\mathcal{C}}}
(g_ag_b)^{-1/2}
\rho^{(ab)}(P_I)
\overline{\prod_{v^{(abc)}_{IJK} \in \Gamma_{\mathcal{C}}} B^{(abc)}_{IJK}
}.
\end{align}

By considering the bulk dual of the annulus partition function, which is the trace over the open Hilbert space, the spectral density is given by
\begin{align}\label{eq:spectrum}
    \rho^{(ab)}(P_I) =g_ag_b\rho_0(P_I)+\delta_{ab} \delta(P_I-\mathbbm{1}).
\end{align}
The first term comes from the open analog of the BTZ saddle, which is the $\mathbb{Z}_2$ quotient of the Euclidean BTZ manifold such that the fixed points are the union of two disks \cite{Takayanagi:2011zk,Fujita:2011fp,Jafferis:2025jle}; the second term comes from the open analog of the thermal AdS saddle, which is the $\mathbb{Z}_2$ quotient of the thermal AdS manifold such that the fixed points are an annulus. The thermal saddle only exists for $a=b$, as the EOW brane is smooth and has only one connected component. States below the black hole threshold can also be included, as in \cite{Miyaji:2022dna,Wang:2025bcx}, but we will postpone their inclusion until Section~\ref{ssec:light}. Incidentally, there are no on-shell wormholes connecting empty disks, so the $g$-functions appearing in \eqref{eq:spectrum} and \eqref{eq:pullout} are constants \cite{Wang:2025bcx}.

\subsection{Action: finite boundary}\label{ssec:actfinite}

In the closed sector, the OPE statistics for scalar above-threshold states can be computed as path integrals for compact regions with fixed-length boundary conditions \cite{Hartman:2025ula}. We now extend this to the open sector. 

The dictionary for compact regions works for non-identity states. However, for computing \eqref{eq:Zexpand}, this is sufficient. If the identity state appears in at least one of the edges in $\Gamma_\mathcal{C}$, we take $B^{(aba)}_{IJ\id}$ outside of the average. In other words, we can focus on the moments of $B_{IJK}^{(abc)}$ where $I,J,K$ are all black hole states. 
The relevant expression then simplifies to
\begin{align}\label{eq:simplify}
\prod_{e^{(ab)}_I \in \Gamma_{\mathcal{C}}}
(g_ag_b)^{-1/2}
\rho^{(ab)}(P_I)\overline{\prod_{v^{(abc)}_{IJK} \in \Gamma_{\mathcal{C}}} B^{(abc)}_{IJK}
}\longrightarrow
\prod_{e^{(ab)}_I \in \Gamma}
(g_ag_b)^{1/2}
\rho_0(I)\overline{\prod_{v^{(abc)}_{IJK} \in \Gamma} B^{(abc)}_{IJK}
},
\end{align}
where $\Gamma$ is obtained from $\Gamma_{\mathcal C}$ by removing all edges labeled by the identity along with any resulting loops, and we have used \eqref{eq:spectrum}.

In terms of the partition function \eqref{eq:Zexpand}, this amounts to pinching off the geodesic intervals at the locations of all identity state cuts, which brings us back to the expression \eqref{eq:Zprod}. The channel decomposition for each $Z[\Sigma_{g_i,n_i}]$ now involves only conformal blocks with no identity lines. 

In the example of \eqref{eq:Sigma04example}, when $M^{(aa)}=\id$, the graph reduces to a $\Theta$ graph, corresponding to
\begin{align}\label{eq:B2eg}
    \overline{B^{(abc)}_{IJK}
    B^{(bac)}_{IKJ}}=\overline{|B^{(abc)}_{IJK}|^2},
\end{align}
where we have used cyclic symmetry and reflection positivity (see e.g.~\cite{Numasawa:2022cni,Hung:2025vgs,Jafferis:2025yxt}).

To study the dictionary for a compact region, the key idea is to remove the conformal blocks from the expression \eqref{eq:Zexpand} so that we can focus on the OPE statistics. Since we are working with the RHS of \eqref{eq:simplify}, the asymptotic boundary $A$ can contain more than one connected component, $A=\sqcup_i A_i$. For each component of the asymptotic boundary $A_i$, we remove the trumpet with topology $A_i\times I$, which keeps the topology of $M$ unchanged. The combination of \eqref{eq:simplify} and the removal of the trumpets eliminates the conformal blocks. 

\begin{figure}
    \centering
    \includegraphics[width=0.4\linewidth]{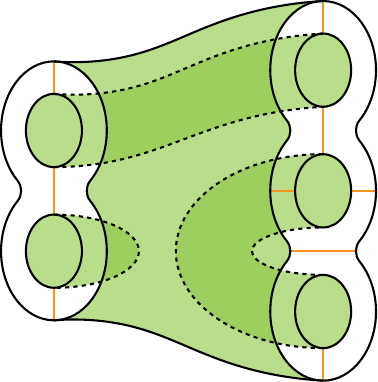}
    \caption{An example of a wormhole contributing to the OPE statistics involving six $B$'s with all operators above the black hole threshold. The left and right boundaries (white) are bordered pleated surfaces $P$, marked with intervals (orange) where the manifold $M$ has corners, and the surfaces connecting them (green) are EOW branes $Q$ which join the boundaries of $A$, i.e., $\partial Q$ = $\partial A$. The surfaces meet perpendicularly at these joints (shown as black circles). (This example is to illustrate a manifold that is allowed kinematically by the rules, but it does not necessarily admit a saddle.)
    }\label{fig:B6eg}
\end{figure}

What replaces the asymptotic boundary $A$ is now a finite surface, which we will call $P$. Figure~\ref{fig:B6eg} shows an example that computes an OPE moment involving six $B_{IJK}^{(abc)}$'s, where all the weights are above the black hole threshold. The fact that $P$ has two connected components means that $\Gamma$ has two graph components, and this could come from a connected $\Gamma_\mathcal{C}$ with some lines set to the identity, according to \eqref{eq:simplify}. The way $\Gamma$ is encoded geometrically is by decomposing $P$ into open pairs of pants and placing corners where the pairs join. The surface $P$ is itself a constant negative curvature 2d manifold, but the extrinsic curvature is singular at the corners, which are geodesic intervals. It is therefore called a pleated surface.

We now present the action relevant for the compact region. Similar to the closed case \cite{Hartman:2025ula}, first define 
\begin{align}
    I_0 =
    I_{\rm bulk} +  I_{\rm brane}+I_{\rm pleated}.
\end{align}
As mentioned earlier, the trace of the extrinsic curvature $K$ is not continuous on $P$ and receives local contributions from the corners $C$. Separating out the smooth part and writing the corner contribution as a 1d integral, which (in general dimensions) is also known as the Hayward term \cite{Hayward:1993my}, we have
\begin{align}
I_{\rm pleated}&=-\frac{1}{8 \pi G_{N}} \int_{P\setminus C} \sqrt{h}\,K
+I_{\rm Hayward},\\
I_{\rm Hayward}&=-
\frac{1}{8 \pi G_N}\int_{C} \sqrt{\gamma}\left(\pi-\theta\right),
\end{align}
where $\theta$ is the angle at the corner as viewed from inside, and $\gamma$ is the induced metric at the codimension-two corner. The Hayward term vanishes when $\theta=\pi$, i.e., when there is no corner.

If one would like to fix the angles at these corners, we can ensure a good variational principle by subtracting off the Hayward term:
\begin{align}
    I_A &= I_0 - I_{\rm Hayward}\nonumber\\
    &=I_{\rm bulk} -\sum_a\chi_a\log g_a-\frac{1}{8 \pi G_{N}} \int_{Q} \sqrt{h}\,K-\frac{1}{8 \pi G_{N}} \int_{ P\setminus C} \sqrt{h}\,K.
\end{align}
The fixed-angle action $I_A$ is therefore simply an integral over the smooth parts of the pleated surface $P\backslash C$ and the EOW brane, plus a topological term for each brane labeled by $a$, generally with different tensions $T_a$ or $g$-functions $g_a$.

To fix the lengths, $I_0$ by itself already ensures a good variational principle, but it is convenient to remove the constant piece in the Hayward term, which makes the action match the semiclassical limit of the Virasoro TQFT partition function \cite{Hartman:2025ula}: 
\begin{align}
    I_L &= I_0 +\frac{1}{8\pi G_N}\int_C\sqrt{\gamma}\,\pi
    \nonumber\\
    &=I_A+\frac{c}{12\pi} \sum_i \ell_i \theta_i,
\end{align}
where $\ell_i$ is the proper length of the $i$-th corner, and $\theta_i$ is the corresponding angle.

Revisiting the example \eqref{eq:B2eg}, the simplest contribution comes from the following topology:
\begin{align}
   \vcenter{\hbox{\includegraphics[height=3cm]{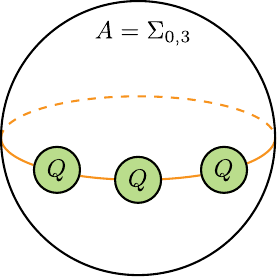}}}~,
\end{align}
where $M=B^3$, $A=\Sigma_{0,3}$ and $Q=D^2\sqcup D^2\sqcup D^2$. Doubling this manifold along the EOW branes gives the closed analog presented in \cite{Hartman:2025ula}. Solving for the saddle, it turns out that the saddle has zero volume, and the corners have angles $\theta_i=0$. Therefore, at least semiclassically,
\begin{align}
    I_L&=I_A=-\log g_a-\log g_b-\log g_c,
    \\
    Z_L&=Z_A=g_ag_bg_c.
\end{align}

By analogy with the closed case \cite{Hartman:2025ula}, we propose that
\begin{align}\label{eq:OPEdict}
\prod_{e^{(ab)}_{I}\in \Gamma}(g_ag_b)^{1/2} \overline{\prod_{v^{(abc)}_{IJK} \in \Gamma} \hat{B}^{(abc)}_{IJK}
}=\sum_{M} Z_L[M, \gamma(\mathbf{P})],
\end{align}
where we have defined the rescaled OPE coefficient
\begin{align}
    \hat{B}^{(abc)}_{IJK}\equiv \frac{B^{(abc)}_{IJK}}{\sqrt{C_0(IJK)}}.
\end{align}
Here, the sum is over all manifolds $M$ with $\partial M=Q\cup P$. The graph $\Gamma$ specifies the decomposition of $P$ into open pairs of pants separated by the intervals $\gamma$ with lengths given by
\begin{align}\label{eq:lengthP}
    \ell_I = 2\pi b P_I.
\end{align}

In the example above, which computes \eqref{eq:B2eg}, the graph has three edges, with state labels $I^{(ab)}$, $J^{(bc)}$ and $K^{(ca)}$. The associated factor of $g$-functions on the LHS of \eqref{eq:OPEdict} is therefore $(g_ag_b)^{1/2}(g_bg_c)^{1/2}(g_cg_a)^{1/2}$. This is precisely reproduced by the presence of three disk branes, contributing $g_a$, $g_b$, and $g_c$, respectively. The relation \eqref{eq:OPEdict} therefore says
\begin{align}
\overline{|\hat{B}^{(abc)}_{IJK}|^2
}\supset 1.
\end{align}
This reproduces results obtained using the dictionary with asymptotic boundaries \cite{Wang:2025bcx,Hung:2025vgs,Jafferis:2025yxt} and the universal expression obtained from bootstrap \cite{Kusuki:2021gpt,Numasawa:2022cni}.

\subsection{Open Virasoro TQFT}\label{ssec:openV}

The path integral with fixed moduli at the asymptotic boundary is related to the fixed-$P$ path integral via
\begin{align}
Z_{\mathrm{oVir}}[M ; \Omega]=\int_0^{\infty} \mathrm{d}^{6g+3n-6} P
\,
\rho_{g,n}^{\mathcal{C}}(\mathbf{P})
\,
Z_{\mathrm{oVir}}^{\mathcal{C}}[M_E, \mathbf{P}] 
\,
\mathcal{F}_{g,n}^{\mathcal{C}}(\mathbf{P} ; \Omega),
\end{align}
where $\rho_{g,n}^{\mathcal{C}}(\mathbf{P})$ was defined in \eqref{eq:rhoC}. The LHS is the path integral on $M$, with fixed moduli $\Omega$ at the asymptotic boundary. The full open-closed TQFT is defined using the Moore-Seiberg conditions \cite{Moore:1988uz,Moore:1988qv} for BCFT, of which there are six \cite{Cardy:1991tv,Lewellen:1991tb}. Among the six, the closed sector enters five of them, rendering them unneeded for the purely open bootstrap. The open Virasoro TQFT can then be built from the remaining condition, namely the crossing symmetry of four boundary operators on the disk \cite{Jafferis:2025yxt}.

The fixed-$P$ partition function $Z_{\mathrm{oVir}}^{\mathcal{C}}[M_E; \mathbf{P}]$ is that of a trivalent Wilson graph $\Gamma$ embedded in $\partial M_E\equiv Q_E$. The embedding manifold $M_E$ is obtained by gluing the asymptotic boundary $A$ to an open handlebody with a boundary Wilson line network specifying the channel $\mathcal{C}$. The trivalent junction is normalized as in \cite{Collier:2023fwi} (via the doubling trick), i.e., multiplying by $C_0(IJK)$ turns the junction into an asymptotic disk.

The building block of open Virasoro TQFT is the open $6j$ manifold:
\begin{align}\label{eq:open6j}
    \vcenter{\hbox{\includegraphics[height=4.5cm]{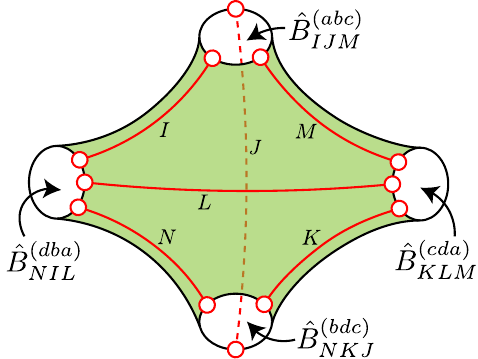}}}
=
    ({g_ag_cg_bg_d})^{-1/2}
    \begin{Bmatrix}
        N & L & I \\
        M & J & K
    \end{Bmatrix}.
\end{align}
The $6j$ symbol is normalized in the Racah-Wigner convention \cite{Teschner:2012em,Teschner:2013tqy,Eberhardt:2023mrq}:
\begin{align}
\left\{\begin{array}{lll}
N & L & I \\
M & J & K
\end{array}\right\}=\frac{1}{\rho_0(K)} \sqrt{\frac{C_0(NLI) C_0(MJ I)}{C_0(L M K) C_0(NJ K)}} \,\mathbb{F}_{IK}\!
\begin{bmatrix}
        M & L \\
        J & N
    \end{bmatrix},
\end{align}
where $\mathbb{F}$ is the Virasoro fusion kernel \cite{Ponsot:1999uf,Ponsot:2000mt}.

The green faces are EOW branes, and the white disks are TQFT state cuts where manifolds can be glued along. Each disk in this diagram represents a geodesic boundary.\footnote{When gluing them, the partition functions simply multiply, in contrast to \cite{Jafferis:2025yxt} where a propagator is needed in each gluing. From this perspective, one can think of this convention as normalizing the propagator to one.} The convention is that, when viewed from \emph{inside} the manifold, it looks like \eqref{eq:datapics}.
The labels $a,b,c,d$ come from different boundary conditions for the BCFT, so each patch of the brane separated by the Wilson lines has an associated $g$-function, related to the tension via \eqref{eq:gandT}. 
Setting all $g$-functions to one gives the partition function of a chiral Virasoro TQFT on the doubled manifold, which is the manifold obtained by gluing it to its mirror image along the EOW brane.

We can then relate OPE statistics to fixed-$P$ partition functions. Using \eqref{eq:Zexpand}, we have
\begin{align}
\overline{
\prod_{e^{(ab)}_I \in \Gamma_{\mathcal{C}}} 
(g_{a}g_b)^{-1/2}\rho^{(ab)}(P_I) \prod_{v^{(abc)}_{IJK} \in \Gamma_{\mathcal{C}}} B^{(abc)}_{IJK}
} 
\supset
\rho_{g,n}^{\mathcal{C}}(\mathbf{P}) 
\,
Z_{\mathrm{oVir}}^{\mathcal{C}}[M_E,\mathbf{P}].
\end{align}
As explained earlier, we can pinch off the identity states to focus on the black hole states. Making the replacement \eqref{eq:simplify}, we obtain the following relation that involves only black hole states:
\begin{align}
\prod_{e^{(ab)}_I \in \Gamma}
(g_ag_b)^{1/2}
\rho_0(I)\overline{\prod_{v^{(abc)}_{IJK} \in \Gamma} B^{(abc)}_{IJK}}
\supset \prod_{e^{(ab)}_I \in \Gamma}
\rho_0(I)\prod_{v^{(abc)}_{IJK} \in \Gamma} C_0({IJK})
\,
{Z_{\mathrm{oVir}}[M_E,\Gamma(\mathbf{P})]}.
\end{align}
Note that we have dropped the label $\mathcal{C}$. Instead, we now use the trivalent graph $\Gamma$, which can have multiple components, to specify the contraction of OPE indices. 

It is useful to define
\begin{align}
    \hat{Z}_{\mathrm{oVir}}[M_E,\Gamma(\mathbf{P})]=Z_{\mathrm{oVir}}
    [M_E,\Gamma(\mathbf{P})]\prod_{v^{(abc)}_{IJK} \in \Gamma} \sqrt{C_0({IJK})}.
\end{align}
This can be thought of as just a choice of the normalization, which is the choice made in \cite{Hartman:2025cyj,Hartman:2025ula}, but it is in some sense more than that: semiclassically, the normalization for $\hat{Z}_{\mathrm{oVir}}^{{\Gamma}}[M_E;\mathbf{P}]$ is such that each trivalent junction is actually a disk boundary with Neumann boundary conditions. We will see in Section~\ref{sec:oc} the utilities of this observation. 

From \eqref{eq:OPEdict}, we therefor identify
\begin{align}
Z_L[M,\gamma(\mathbf{P})]=
\hat{Z}_{\mathrm{oVir}}[M_E,\Gamma(\mathbf{P})].
\end{align}
The analogous relation in the closed sector is given in \cite{Hartman:2025ula}, and reviewed in a later section in \eqref{eq:closedZL}. 

Consider $M_E=B^3$ so that $Q_E\equiv\partial M_E=S^2$. We can then draw trivalent Wilson graphs $\Gamma$ on $Q_E$. Two such examples are given by
\begin{align}\label{eq:Vgrapheg}
    \vcenter{\hbox{\includegraphics[height=3.5cm]{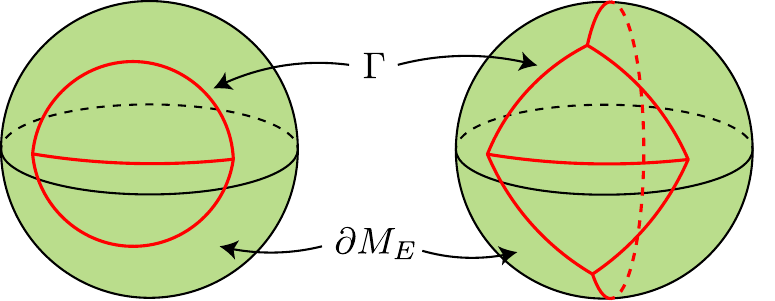}}}~,
\end{align}
which are respectively known as the $\Theta$ graph and the $K_4$ graph.

\section{Triangulation}\label{sec:trig}

A triangulation in three dimensions is a way of decomposing a 3-manifold into tetrahedra. It plays the same role that triangulations by triangles play for two-dimensional surfaces, but the combinatorics and topology are much richer in one higher dimension. Triangulation encodes the global topology of the space in discrete data: which tetrahedra are present, how their faces are identified, and how edges and vertices fit together. This makes triangulations extremely useful in practice.

\subsection{Semiclassical tetrahedral decomposition}\label{ssec:semi}

Saddles contributing to the OPE moments can be constructed via tetrahedral decomposition. The idea is to build the manifold by gluing together copies of the following object:
\begin{align}\label{eq:tet_allmild}
        \vcenter{\hbox{\includegraphics[height=3cm]{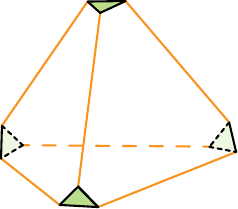}}}~.
\end{align}
This is an example of a generalized hyperbolic tetrahedron \cite{ThurstonBook,ushijima2006volume}. It is a tetrahedron with all four vertices ``chopped off'', so we will also refer to it as the truncated tetrahedron. 

Each truncated tetrahedron has four hexagonal faces and four triangular ones. 
The triangular faces and hexagonal faces are orthogonal to each other, i.e., they always intersect at an angle of $\pi/2$. All eight faces have zero extrinsic curvature. We have colored the triangular faces green to represent EOW branes and will refer to them as \emph{EOW faces}; we will refer to the uncolored faces as \emph{OPE faces}. 

To construct saddles, we glue along OPE faces, but not EOW faces. For a valid gluing, the two faces must have exactly the same shape. This is ensured by having the edge lengths match pairwise: for truncated tetrahedra, if the three lengths of a face are identified with the three lengths of another face (which could be of the same tetrahedron or a different one) in an orientable way (the normals should be opposite to each other), then the two faces are glued smoothly. 

Some of the edges become fully surrounded, i.e., they are not part of $\partial M$ in the resulting manifold. These edges are called \emph{internal}. Smoothness at an internal edge requires that there is no conical deficit or excess, meaning that the dihedral angles around it add up to $2\pi$.

As the hexagonal faces are perpendicular to the triangular faces, gluing ensures that the triangular faces join smoothly. The unglued hexagonal faces form pleated surfaces. 
The resulting manifold will therefore satisfy the boundary conditions specified in Section~\ref{ssec:actfinite}. 
Together, the boundary of the resulting manifold is
\begin{align}
    \partial M= Q \cup P.
\end{align}

In the open case, it can sometimes be easier to construct examples than the closed case studied in \cite{Hartman:2025ula}. For example, even without gluing, the truncated tetrahedron is a valid configuration on its own. In fact, it is the same as the open VTQFT diagram on the RHS of \eqref{eq:Vgrapheg}. The unglued OPE faces form a surface corresponding to
\begin{align}
    g_ag_bg_cg_d\,\overline{\hat{B}^{(abc)}_{IJK}\hat{B}^{(cda)}_{KLM}\hat{B}^{(bdc)}_{NKJ}\hat{B}^{(dba)}_{NIL}}.
\end{align}
Each triangular EOW contributes a factor of $g$, reproducing the correct $g$-function dependence.

Consider a more complicated example, where four tetrahedra are glued together:
\begin{align}\label{eq:gluet}
    \vcenter{\hbox{\includegraphics[height=6cm]{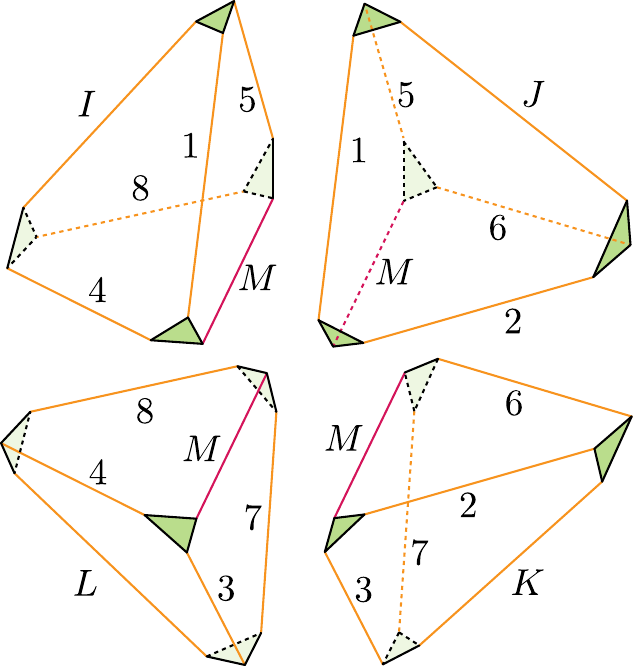}}}~\longrightarrow~
    \vcenter{\hbox{\includegraphics[height=6cm]{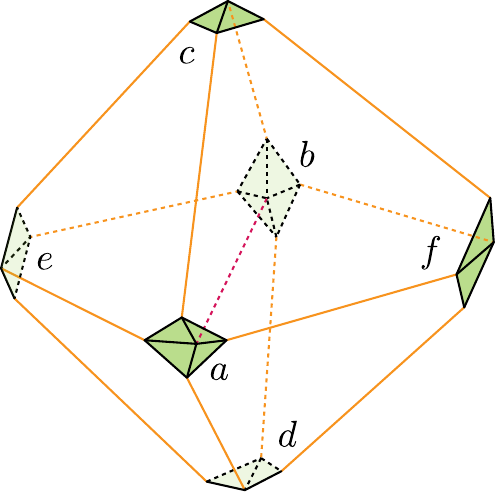}}}
    ~.
\end{align}
The resulting manifold has 8 unglued OPE faces, and the EOW faces join to form 6 connected components. Despite the drawing, the EOW faces join smoothly with no kinks, as a result of the truncated tetrahedron having all of its faces orthogonal to each other at the edges. The points where four EOW faces join on the front and back two components of the EOW brane (labeled $a$ and $b$) are connected by an internal edge (dark pink). Requiring that the total angle around it be $2\pi$ fixes its length as a function of all the external lengths (there are 12 of them). 

To solve for the saddle, it is useful to organize the data in terms of matrices \cite{ThurstonBook,ushijima2006volume}. See \cite{Hartman:2025ula} for a recent review. Define the length Gram matrix
\begin{align}
\tilde{G}_{i j}\equiv -\cosh \ell_{i j},
\end{align}
where $\ell_{ij}$ is the length between vertices $i$ and $j$, and the angle Gram matrix
\begin{align}\label{eq:Gij}
    G_{ij}\equiv-\cos\psi_{ij}, 
\end{align}
where $\psi_{ij}$ is the dihedral angle between the faces opposite to vertices $i$ and $j$. They are related by
\begin{align}\label{eq:GGtil}
    \tilde{G}=M G^{-1} M,\quad M_{ij}\equiv\frac{\delta_{ij}}{\sqrt{G_{ii}^{-1}}}.
\end{align}
It is convenient to use the following convention for drawing a tetrahedron:
\begin{align}
    \vcenter{\hbox{\includegraphics[height=3cm]{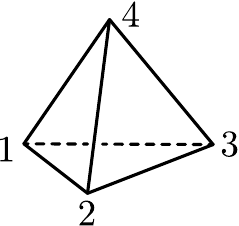}}}~\longrightarrow~\vcenter{\hbox{\includegraphics[height=2.5cm]{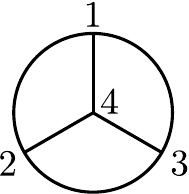}}}.
\end{align}
This gives an ordering to the vertices so that the rows and columns of the Gram matrices are ordered according to it. With this, the four tetrahedra in \eqref{eq:gluet} can be represented by
\begin{align}
    \vcenter{\hbox{\includegraphics[height=2cm]{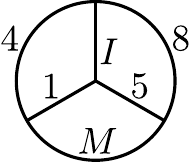}}}~,\quad
    \vcenter{\hbox{\includegraphics[height=2cm]{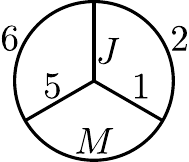}}}~,\quad
    \vcenter{\hbox{\includegraphics[height=2cm]{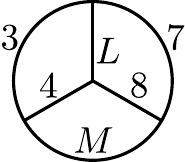}}}~,\quad
    \vcenter{\hbox{\includegraphics[height=2cm]{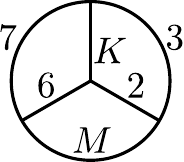}}}~.
\end{align}
The length Gram matrices are then given by
\begin{align}
\tilde{G}^{(1)}&=\left(\begin{array}{cccc}
1 & -\cosh \ell_4 & -\cosh \ell_8 & -\cosh \ell_I \\
-\cosh \ell_4 & 1 & -\cosh \ell_M & -\cosh \ell_1 \\
-\cosh \ell_8 & -\cosh \ell_M & 1 & -\cosh \ell_5 \\
-\cosh \ell_I & -\cosh \ell_1 & -\cosh \ell_5 & 1
\end{array}\right),
\\
\tilde{G}^{(2)}&=\left(\begin{array}{cccc}
1 & -\cosh \ell_6 & -\cosh \ell_2 & -\cosh \ell_J \\
-\cosh \ell_6 & 1 & -\cosh \ell_M & -\cosh \ell_5 \\
-\cosh \ell_2 & -\cosh \ell_M & 1 & -\cosh \ell_1\\
-\cosh \ell_J & -\cosh \ell_5 & -\cosh \ell_1 & 1
\end{array}\right),
\\
\tilde{G}^{(3)}&=\left(\begin{array}{cccc}
1 & -\cosh \ell_3 & -\cosh \ell_7 & -\cosh \ell_L 
\\
-\cosh \ell_3 & 1 & -\cosh \ell_M & -\cosh \ell_4 
\\
-\cosh \ell_7 & -\cosh \ell_M & 1 & -\cosh \ell_8
\\
-\cosh \ell_L & -\cosh \ell_4 & -\cosh \ell_8 & 1
\end{array}\right),
\\
\tilde{G}^{(4)}&=\left(\begin{array}{cccc}
1 & -\cosh \ell_7 & -\cosh \ell_3 & -\cosh \ell_K 
\\
-\cosh \ell_7 & 1 & -\cosh \ell_M & -\cosh \ell_6 
\\
-\cosh \ell_3 & -\cosh \ell_M & 1 & -\cosh \ell_2
\\
-\cosh \ell_K & -\cosh \ell_6 & -\cosh \ell_2 & 1
\end{array}\right).
\end{align}
Notice that we have already used most of the gluing conditions, which is that some lengths are identified. (Otherwise all the 24 lengths would be independent.) What remains is to ensure smoothness around the internal edges. In this case, there is only one, and the condition is given by
\begin{align}
    \psi^{(1)}_{14}+\psi^{(2)}_{14}+\psi^{(3)}_{14}+\psi^{(4)}_{14}=2\pi.
\end{align}
With \eqref{eq:Gij} and \eqref{eq:GGtil}, this allows one to solve for $\ell_M$, as all the other lengths are given as input (boundary conditions). For example, when all external lengths are taken to be $1.000$, we numerically obtain $\ell_M=4.086$. Once $\ell_M$ is known, the volume of each tetrahedron can be computed, and the volume of the manifold $M$ is simply the sum of their volumes. The formula for the volume of a truncated tetrahedron \cite{ushijima2006volume} in terms of the six lengths can be obtained as the classical limit of the $6j$ symbol \cite{Teschner:2012em,Chen:2024unp,Hartman:2025ula,Liu:2025inv,Liu:2025tzv}, with the length related to $P$ via \eqref{eq:lengthP}. In this example, with all external lengths set to $1.000$, the volume is computed to be $9.276$.

\subsection{Fixed-length ensemble from open Virasoro TQFT}

At the quantum level, we can build manifolds using the open $6j$ manifold \eqref{eq:open6j}. The open $6j$ manifolds are glued along the disks, and when the boundary Wilson lines form a closed loop, one performs the annular surgery:
\begin{align}
    \int \d P\, g_ag_b\rho_0(P)\,
    \vcenter{\hbox{\includegraphics[height=2cm]{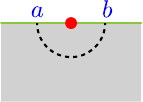}}}\times S^1 = 
    \vcenter{\hbox{\includegraphics[height=2cm]{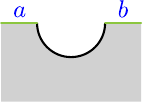}}}\times S^1
    \cup
    \vcenter{\hbox{\includegraphics[height=2.5cm]{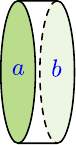}}}~,
\end{align}
where the annuli are glued to each other on the RHS of the equation. To be more precise, we remove a neighborhood of the Wilson loop, leaving a cut with annulus topology, and glue to it a slab, with topology $S^2\times I$. The slab has two disk EOW branes, generally with different $g$-functions. 

As an example, consider gluing four open $6j$ manifolds in the following way:
\begin{align}\label{eq:glue1}
    \vcenter{\hbox{\includegraphics[height=5.5cm]{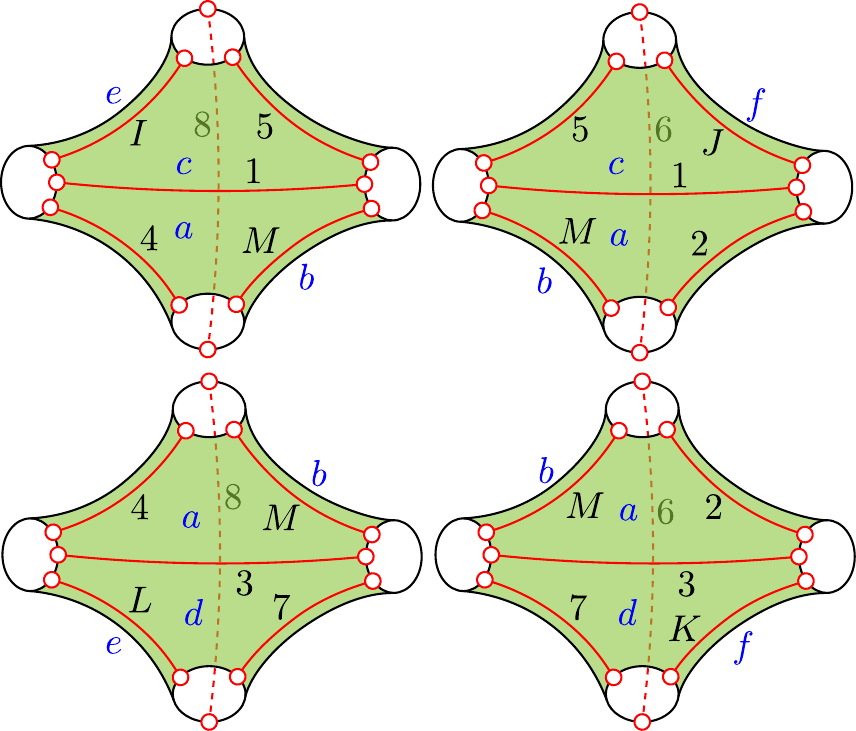}}}=\vcenter{\hbox{\includegraphics[height=5.5cm]{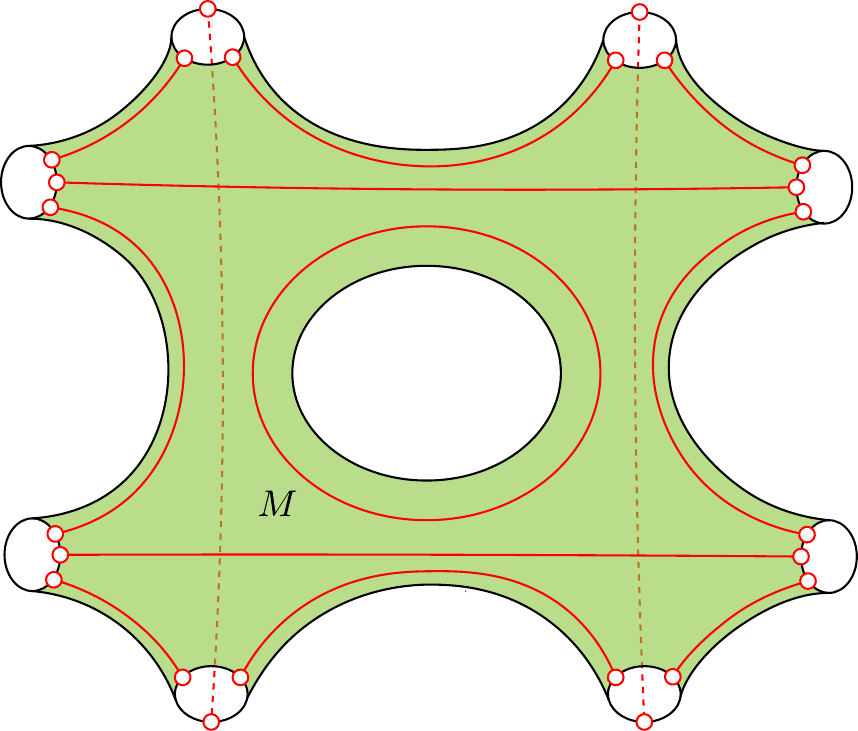}}}.
\end{align}
This forms a Wilson loop, so we perform the annular surgery:
\begin{align}
    &\int \d P_M\,g_ag_b\rho_0(P_M)
    \times \eqref{eq:glue1}
   \nonumber \\
    =&
    \vcenter{\hbox{\includegraphics[height=5.5cm]{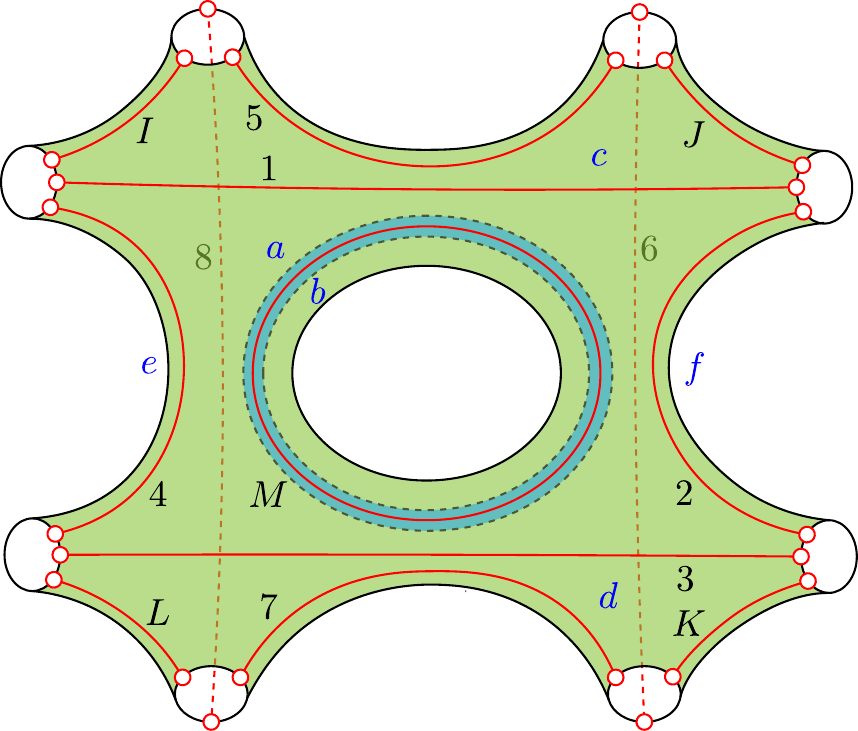}}} \cup
    \vcenter{\hbox{\includegraphics[height=2cm]{figs/slab.pdf}}}
   =\vcenter{\hbox{\includegraphics[height=5cm]{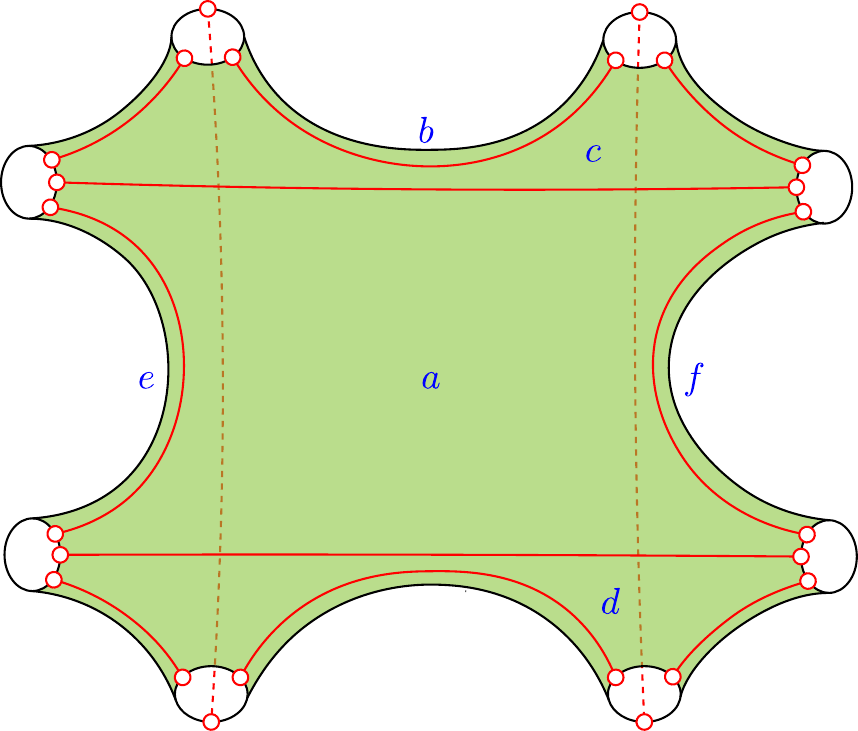}}}.
\end{align}
In this example, the slab removes the genus hole, changing the topology of both the EOW brane and the bulk manifold. Shrinking the disks to points, the corresponding open Virasoro TQFT diagram looks like
\begin{align}
    \vcenter{\hbox{\includegraphics[height=4cm]{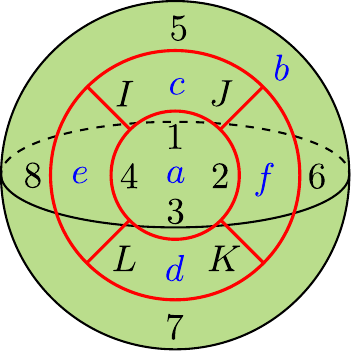}}}
    ~,
\end{align}
where $M_E=B^3$ and $Q_E=S^2$. The graph $\Gamma$ is isomorphic to the graph of the edges of a cube, and hence is the hypercube graph $Q_3$.

Recalling from Section~\ref{ssec:openV} how the graph corresponds to the pleated surface on the boundary of a compact region of spacetime, it is straightforward to see that this example corresponds to \eqref{eq:gluet}. Each of the starting four open $6j$ manifolds corresponds to a truncated tetrahedron, and the annular surgery corresponds to the formation of an internal edge.

The feature demonstrated by this example is general: gluing open $6j$ manifolds and performing annular surgeries is the same as a tetrahedral decomposition, where some hexagonal (OPE) faces are glued and the triangular (EOW) faces are unglued. The OPE faces that are unglued join to become the pleated surface. In other words, each open $6j$ manifold (which is itself a truncated tetrahedron) is associated to a dual truncated tetrahedron as follows:
\begin{align}
    \vcenter{\hbox{\includegraphics[height=4cm]{figs/6jlabeled.pdf}}}
    ~\longrightarrow~
    \vcenter{\hbox{\includegraphics[height=4cm]{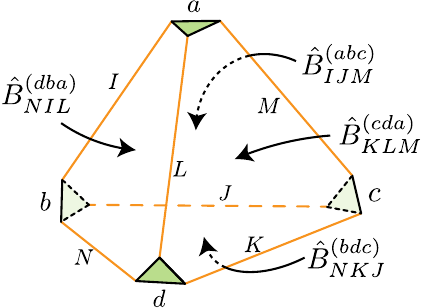}}}~.
\end{align}
Whenever a Wilson loop forms in the tetrahedral decomposition, an internal edge forms in the dual picture.

To summarize, the open Virasoro TQFT partition function can be computed by multiplying the $6j$ symbols associated to each dual tetrahedron and integrating over the weights of the internal edges:
\begin{align}\label{eq:oCTV}
\hat{Z}_{\mathrm{oVir}}[M_E, \Gamma(\mathbf{P})]=\prod_{I}\left(\int_0^{\infty} \d \mathring{P}_I\,g_ag_b \rho_0(\mathring{P}_I)\right) \prod_{\hat{\Delta} \in \hat{T}} \hat{W}(\hat{\Delta}),
\end{align} 
where $\hat{W}(\hat{\Delta})$ is the function \eqref{eq:open6j} associated to each truncated tetrahedron $\hat{\Delta}$ that belongs to a triangulation $\hat{T}$ of the resulting manifold, and the integral is over all internal edges with weights denoted with a ring. The resulting manifold is related to $(M_E,\Gamma(\mathbf{P}))$ via the shrinking of the pleated surface $P$ to the graph $\Gamma$.

It is worth emphasizing that this is now a quantum duality, which reproduces the semiclassical tetrahedral decomposition via saddle-point approximation \cite{Hartman:2025ula}. One may refer to this as the open CTV theory to distinguish it from the more restricted version we discuss later, where no OPE faces are left unglued. 

The convergence of the integral \eqref{eq:oCTV} was studied in \cite{Liu:2025inv}. In particular, when all OPE faces are glued, the resulting manifold has totally geodesic boundary, and the corresponding integrals converge, thereby rendering the TQFT well defined. In this case, each boundary component is a closed surface of genus at least two, assembled from the triangular EOW faces of the truncated tetrahedra.

\subsection{Below-threshold states}\label{ssec:light}

So far, we have considered only black hole states, which are states with real $P$. We can also consider states below the black hole threshold. In the open sector of the BCFT, these are states in the open-string Hilbert space with $0<h<(c-1)/24$, so that $P$ is imaginary. The bulk dual of such an operator is a particle that is constrained to stay on the EOW brane. Geometrically, they look like the corners that we introduced in earlier sections, but we will refer to these as \emph{kinks} to conceptually distinguish them from the corners. 

The actions change only slightly with the inclusion of kinks. It is still given by \eqref{eq:action_tot}, except that $I_{\rm brane}$ should integrate over only the smooth part of $Q$, i.e., over $Q\backslash K$, where $K$ is the kink \cite{Wang:2025bcx}. It has a good variational principle when the angle is fixed. It reduces to \eqref{eq:action_tot} when there are no kinks.

\begin{figure}
    \centering
    \includegraphics[width=0.95\linewidth]{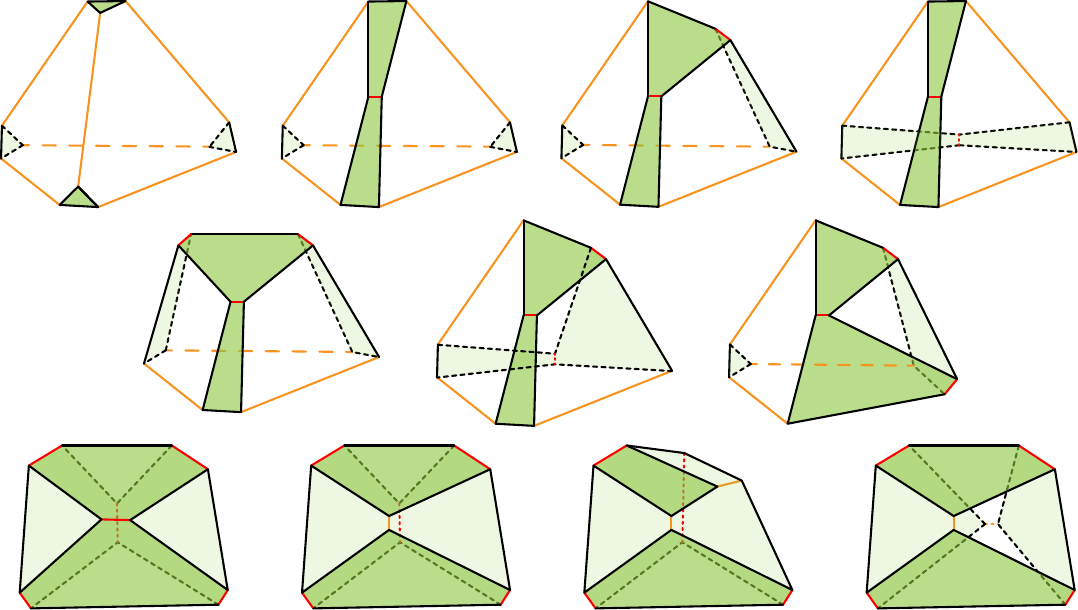}
    \caption{A list of all possible shapes of hyperbolic tetrahedra with hyperideal vertices. The colored faces (green) are EOW faces, whereas the uncolored ones are OPE faces. Each OPE face is associated to an OPE coefficient $\hat{B}_{IJK}^{(abc)}$, which is also an open pair of pants. When an edge goes below the threshold, the EOW branes on the two ends of the edge become connected, e.g., when going from the first to the second tetrahedron. Once this happens, the edge becomes dualized, and the dual edge (now red) is called a kink, where the normal of the EOW brane changes discontinuously. 
    The ones in the third row are the same as those in the first row, except that the EOW brane and OPE faces are swapped.}
    \label{fig:tetraall}
\end{figure}

In the presence of below-threshold states, the building blocks for the semiclassical tetrahedral decomposition explained in Section~\ref{ssec:semi} are modified. Illustrations of all possible building blocks are given in Figure~\ref{fig:tetraall}. Each time a black hole state is analytically continued below the threshold, a corner (orange) is dualized to a kink (red), and the EOW branes (green) become joined at the kink:
\begin{align}
    \vcenter{\hbox{\includegraphics[height=2.5cm]{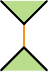}}}
    ~\longrightarrow~
    \vcenter{\hbox{\includegraphics[height=2cm]{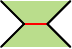}}}~.
\end{align}

In terms of the partition function, \eqref{eq:oCTV} still holds, but the external weights $\mathbf{P}$ that correspond to below-threshold states are analytically continued to imaginary values. The boundary conditions corresponding to $\hat{Z}_{\rm oVir}[M_E,\Gamma(\mathbf{P})]$ are such that we fix
\begin{equation}
\begin{cases}
\ell_I=2\pi bP_I,& \text{above-threshold};
    \\\psi_I=2 \pi \i b P_I,&\text{below-threshold}.
\end{cases}
\end{equation}

\section{An open-closed duality}\label{sec:oc}

\subsection{Relation between CTV and closed Virasoro TQFT}
There is a relation between two copies of Virasoro TQFT and CTV \cite{Hartman:2025cyj}:
\begin{align}\label{eq:CTVVir}
Z_{\mathrm{CTV}}[\tilde{M}_E, \tilde{\Gamma}(\mathbf{P})]=\prod_{i}\left(\int_0^{\infty} \d P_i' \,\mathbb{S}_{P_i P_i^{\prime}}[\id]\right)\left|\hat{Z}_{\mathrm{Vir}}[\tilde{M}_E, \tilde{\Gamma}(\mathbf{P}^{\prime})]\right|^2.
\end{align}
The partition function for two copies of Virasoro TQFT is in the closed sector of the open-closed TQFT, so the embedding manifold $\tilde{M}_E$ has no boundary, and $\tilde{\Gamma}$ is a graph of bulk Wilson lines. Furthermore, $P'=\bar{P}'$ on the RHS, so it is further restricted to the scalar sector of the closed sector. The tildes indicate ``closed'', and the tilded objects are not superficially related to the untilded ones.

The CTV partition function is defined by
\begin{align}
{Z}_{\rm CTV}[\tilde{M}_E, \tilde{\Gamma}(\mathbf{P})]=\prod_{j}\left(\int_0^{\infty} \d \mathring{P}_j\, \rho_0(\mathring{P}_j)\right) \prod_{\Delta \in T} W(\Delta),
\end{align} 
where $\mathring{P}_j$ denotes the weight of internal edges, $T$ is a large triangulation of $(\tilde{M}_E,\tilde{\Gamma})$ (meaning that there are no internal vertices), and $\Delta$ labels the tetrahedra that form the triangulation, with some edges internal, and others belonging to $\tilde{\Gamma}$. The function $W(\Delta)$ is given by 
\begin{align}
    W\left(\vcenter{\hbox{\includegraphics[height=2cm]{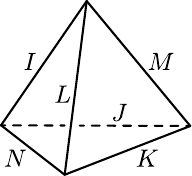}}}\right)=\begin{Bmatrix}
        N & L & I \\
        M & J & K
    \end{Bmatrix}.
\end{align}

The Virasoro TQFT partition function $\hat{Z}$ is related to the one defined in \cite{Collier:2023fwi} by
\begin{align}\label{eq:Zvir2norm}
    |\hat{Z}_{\mathrm{Vir}}[\tilde{M}_E,\tilde{\Gamma}(\mathbf{P})]|^2
    =
    |{Z}_{\mathrm{Vir}}[\tilde{M}_E,\tilde{\Gamma}(\mathbf{P})]|^2
    \prod_{v_{ijk} \in \tilde{\Gamma}} \sqrt{|C_0({ijk})|^2}.
\end{align}
As mentioned earlier, it is useful to take the perspective that this is more than a redefinition. Geometrically, the factor turns each trivalent junction into a Neumann boundary (a three-punctured sphere with zero extrinsic curvature). At the quantum level, the Neumann boundary condition is reflected in the fact that gluing along two such three-punctured spheres requires no additional factors.

\subsection{Derivation}
The classic derivation of \eqref{eq:CTVVir} uses the chain-mail invariant \cite{Roberts1995} by relating both sides to the partition function of a chain-mail link \cite{Barrett:2002vi,Garcia-Islas:2004lwa,Barrett:2004im}, with some changes to avoid singularities in the Virasoro case \cite{Hartman:2025cyj}. We now provide an alternative perspective by deriving it using an open-closed duality. The derivation avoids the mention of the chain-mail link and has rather simple pictorial descriptions. For the purpose of the rest of this section, we set all $g$-functions to one, as the argument we use needs only tensionless branes. 

Start by writing the LHS of \eqref{eq:CTVVir} as a fixed-angle path integral \cite{Hartman:2025ula}:
\begin{align}\label{eq:CTV=A}
Z_{\mathrm{CTV}}[\tilde{M}_E, \tilde{\Gamma}(\mathbf{P})] 
=
\tilde{Z}_A[\tilde{M}, \tilde{\gamma}(\mathbf{P})],
\end{align}
where $\tilde{M}=\tilde{M}_E-N(\tilde{\Gamma})$, i.e., $\tilde{M}$ is obtained by removing the neighborhood $N(\tilde{\Gamma})$ of the graph, and $\tilde{\gamma}$ is the set of circular corners separating $\partial\tilde{M}=\partial N(\tilde{\Gamma})$ into ordinary pairs of pants.

This reads identically to the definition of the open Virasoro TQFT partition function \eqref{eq:oCTV} (as all branes are now tensionless). In fact, it is a special case of \eqref{eq:oCTV} when all OPE faces are glued together:
\begin{align}\label{eq:CTV=oV}
\tilde{Z}_A[\tilde{M}, \tilde{\gamma}(\mathbf{P})]=
\hat{Z}_{\rm oVir}[M_E,\Gamma(\mathbf{P})],
\end{align}
where the ${P}$'s are imaginary, corresponding to fixing corner angles on the LHS and kink angles on the RHS. From Figure~\ref{fig:tetraall}, it is clear that when all OPE faces are glued together, the remaining faces will be EOW branes with kink loops, i.e., $\Gamma$ is not a trivalent graph but a union of loops. More precisely, the boundary $\partial M_E=Q_E$ is a closed genus-$g$ surface separated by $3g-3$ kink loops into ordinary pairs of pants. (Consequently, the hat on the RHS can be removed.) Recall that kinks are geometrically the same as corners, so we identify 
$\Gamma(\mathbf{P})$ with $\tilde{\gamma}(\mathbf{P})$, and $M_E$ with $\tilde{M}$. 

An example of $(M_E,\Gamma)$ is
\begin{align}\label{eq:oc1}
    \vcenter{\hbox{\includegraphics[height=3cm]{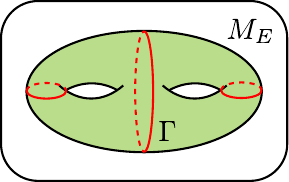}}}~,
\end{align}
where we have drawn a portion of $M_E$ that is ``outside'' of the genus-2 surface as shown.

We are now ready to discuss the main step in the argument, which uses the open-closed duality. The open-closed duality in BCFT is the statement that the annulus partition function can be computed in two ways: in an open-string channel or in a closed-string channel
\cite{Cardy:1991tv,Lewellen:1991tb}. In the $P$ basis, the two channels are related by
\begin{align}\label{eq:ocdual}
    \vcenter{\hbox{\includegraphics[height=3cm]{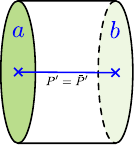}}}~=\int_0^{\infty} \d P\, \mathbb{S}_{P'P}[\id]~
    \vcenter{\hbox{\includegraphics[height=3cm]{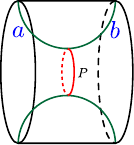}}}~,
\end{align}
where
\begin{align}
    \mathbb{S}_{P_1P_2}[\id]=2\sqrt{2}\cos(2\pi P_1P_2)
\end{align}
is the modular $S$ crossing kernel on a torus. This relation follows via the doubling trick, which turns the annulus to a torus \cite{Cardy:1984bb,Cardy:1991tv,Numasawa:2022cni}. The LHS is a solid cylinder with two disk EOW branes punctured by the bulk Wilson line (blue) extending between them. It is the 3d path integral that prepares a fixed-$P$ closed-string state propagating from one border to the other. It can only be a scalar, as the one-point function of a bulk operator on the disk is zero for spinning insertions (by conservation of angular momentum). The RHS is a solid torus (or half of it) whose boundary is composed of the annulus where the state lives and an annular EOW brane with a boundary Wilson loop (red) running along the circle direction, separating the brane into two annular patches, each with a generally different $g$-function. (It is the product of a half-disk and a circle.) It is the 3d path integral that prepares a fixed-$P$ open-string state running in a loop.

Let us now apply this to \eqref{eq:CTV=oV} and see what happens. Let us illustrate it with the example \eqref{eq:oc1}. From the open-closed duality \eqref{eq:ocdual}, doing the S transforms on each of the three closed loops gives
\begin{align}\label{eq:cVgraph}
    \vcenter{\hbox{\includegraphics[height=3cm]{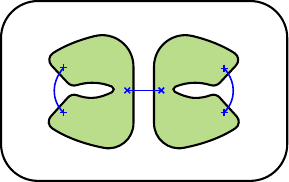}}}~=~\vcenter{\hbox{\includegraphics[height=3cm]{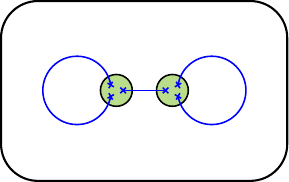}}}~.
\end{align}
The EOW brane changes from a genus-2 surface to the union of two spheres, each with three punctures. The punctures are connected by three scalar bulk Wilson lines. Now, we realize the importance of the normalization \eqref{eq:Zvir2norm}: $|\hat{Z}_{\rm Vir}|^2$ is the partition function when the trivalent junctions are replaced with Neumann spheres. Recall that the EOW branes are indeed Neumann boundaries, so the RHS of this equation is exactly the hatted partition function $|\hat{Z}_{\rm Vir}|^2$ on $(\tilde{M}_E,\tilde{\Gamma})$, where $\tilde{\Gamma}$ is the scalar bulk Wilson graph obtained by shrinking the spheres to points, and $\tilde{M}_E$ is the embedding manifold for this graph: 
\begin{align}
    \vcenter{\hbox{\includegraphics[height=3cm]{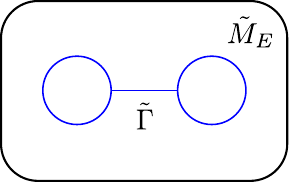}}}~.
\end{align}

This example straightforwardly generalizes to any $(M_E,\Gamma)$ where $\Gamma$ is a union of loops that separate $\partial M_E=Q_E$ into ordinary pairs of pants. The general statement is therefore
\begin{align}\label{eq:ViroVir}
\left|\hat{Z}_{\mathrm{Vir}}[\tilde{M}_E, \tilde{\Gamma}(\mathbf{P}^{\prime})]\right|^2=\prod_{i}\left(\int_0^{\infty} \d P_i \,\mathbb{S}_{P_i^{\prime}P_i }[\id]\right) \hat{Z}_{\rm oVir}[M_E,\Gamma(\mathbf{P})].
\end{align}
This is the main result. Using \eqref{eq:CTV=A}, \eqref{eq:CTV=oV}, and the fact that the S kernel squares to one, we see that we have reproduced \eqref{eq:CTVVir}. 

\subsection{Boundary conditions}

The open Virasoro TQFT computes the open OPE statistics with fixed-length boundary conditions for above-threshold states and fixed-angle boundary conditions for below-threshold states. 

According to \eqref{eq:CTV=oV}, the kinks representing below-threshold open states become corners representing above-threshold closed states. This is the reason behind the fact that CTV computes the fixed-angle ensemble for above-threshold states. 

The S transform in \eqref{eq:ViroVir} plays the role of a Fourier transform (or Laplace transform). Semiclassically, it turns the action from one that has a good variational principle with fixed-angle boundary conditions to one that has fixed-length boundary conditions. Consequently, 
\begin{align}\label{eq:closedZL}
\tilde{Z}_L[\tilde{M},\tilde{\gamma}(\mathbf{P})]=\left|\hat{Z}_{\mathrm{Vir}}[\tilde{M}_E, \tilde{\Gamma}(\mathbf{P})]\right|^2,
\end{align}
where $\tilde{Z}_L$ is path integral with fixed lengths for the marked circles $\tilde{\gamma}$, each corresponding to an edge of the trivalent graph $\tilde{\Gamma}$ \cite{Hartman:2025ula}.

Starting with the closed Virasoro TQFT partition function with a scalar Wilson graph, we could also take some of the weights to be below the threshold. For such states, the bulk Wilson line becomes a conical defect, and the length analytically continues to the conical angle around the defect. The closed Virasoro TQFT therefore computes the closed OPE statistics with fixed-length boundary conditions for above-threshold states and fixed-conical-angle boundary conditions for below-threshold states. For example, in \eqref{eq:cVgraph}, if the middle Wilson line is taken to be below the threshold, the boundary of the corresponding manifold $\tilde{M}=\tilde{M}_E-N(\tilde{\Gamma})$ becomes two punctured tori connected by a conical defect. 

We summarize the different objects in Table~\ref{tab:boundaryconds}. We see from the table that the story is similar for the diagonal entries, while the off-diagonal entry has the opposite behavior. We now understand this from two perspectives. Going from left to right in the first row, the change in the boundary conditions is a result of the Laplace transform; going from bottom to top in the second column, the change is a result of the open-closed duality, where the EOW brane becomes a state (OPE) boundary and the open below-threshold Wilson loops turn into dual circles of the closed above-threshold states.

\begin{table}[t]
\centering
\renewcommand{\arraystretch}{1.3}
\begin{tabular}{c c c c}
\toprule
 & & \textbf{closed TQFT (scalars)} & \textbf{open TQFT} \\
\midrule
\multirow{2}{*}{closed OPE}
& above-threshold & fix length (dual) & fix angle (corner) \\
& below-threshold & fix angle (conical) & fix length \\
\midrule
\multirow{2}{*}{open OPE}
& above-threshold & N/A & fix length (dual) \\
& below-threshold & N/A & fix angle (kink) \\
\bottomrule
\end{tabular}
\caption{Summary of boundary conditions. On the diagonal, the length that is fixed for an above-threshold state (either open or closed) is also known as the dual length because it is the length of the edge that is obtained by dualizing the Wilson line representing the state. In contrast, the length that is fixed for a below-threshold closed state for the CTV partition function computed via the open Virasoro TQFT is the longitudinal length, which is along the direction of the Wilson line (geometrized as a conical defect).}
\label{tab:boundaryconds}
\end{table}

\section{Discussion}\label{sec:disc}

To summarize, we started by setting up a purely open ensemble by restricting ourselves to the open sector of the BCFT, where the only data are the spectrum of boundary operators, their OPE coefficients, and the $g$-functions associated to boundary conditions. This ensemble is already nontrivial and provides a simpler setting in which to study the relation between 3d gravity and 2d CFT. For instance, there are no modular symmetry constraints for the spectral density, and the contributing manifolds are significantly fewer. We also find no evidence for the existence of accumulation points in the set of contributing manifolds. We then explained how to construct the contributing hyperbolic manifolds by tetrahedral decomposition, which is also equivalent to gluing open $6j$ manifolds and performing annular surgery. We then returned to the full open-closed Virasoro TQFT and used an open-closed duality to explain a relation between two objects: the diagonal sector of two copies of Virasoro TQFT, which may be viewed as the closed sector with scalar bulk Wilson lines, and CTV, which may be viewed as a special class of partition functions of the open sector that involves only boundary Wilson loops.

An interesting application of the purely open model is to study the CFT partition function on a closed Riemann surface \cite{Hung:2019bnq,Brehm:2021wev,Chen:2022wvy,Cheng:2023kxh}. Start with a CFT on $\Sigma_g$ and consider a 2d triangulation of the surface. At each vertex, we then regularize the triangulation by removing its small neighborhood:
\begin{align}\label{eq:trig1}
    \vcenter{\hbox{\includegraphics[height=3cm]{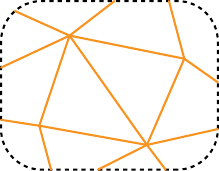}}}~
    \longrightarrow~\vcenter{\hbox{\includegraphics[height=3cm]{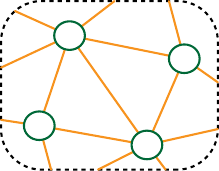}}}~.
\end{align}
This turns it into a bordered Riemann surface $\Sigma_{g,n}$, where $n$ is the number of vertices of the triangulation. The idea is then to compute the BCFT partition function on $\Sigma_{g,n}$. The triangulation we started with has naturally given us an open channel decomposition, where each hexagon is associated to a boundary-to-boundary OPE coefficient $B_{IJK}^{(abc)}$. For rational CFTs, with certain ``shrinkable'' boundary conditions imposed, this approach can be used to reproduce the partition function of the closed CFT in the limit the holes shrink \cite{Hung:2019bnq,Brehm:2021wev,Chen:2022wvy,Cheng:2023kxh,Brehm:2024zun}. It also has an interpretation in the bulk as a sum over geometries \cite{Chen:2024unp,Hung:2024gma,Bao:2024ixc,Hung:2025vgs,Geng:2025efs,Jafferis:2025yxt}. 

For a fixed number of vertices, any triangulation of a 2d surface can be reached from any other via 2-2 Pachner moves. The 2-2 Pachner move translates to the F move on the BCFT channel decomposition. It would be natural to extend the argument of \cite{Belin:2026pko,Wang:2025jgo} to show that the sum over 3d geometries is invariant under such moves. A main difference, however, is that the identity state cannot appear when the boundary conditions are different on either side of the open state cut, so the open analogs of the handlebodies are generally absent. This is a new feature that needs to be taken into account in the argument. Another related subtlety is the counting of $g$ factors: crossing transformations can change the counting, and the topology of the EOW brane must change correspondingly to reproduce this change. To reach a triangulation with a different number of vertices, one also needs 1-3 and 3-1 Pachner moves. It would be interesting to understand whether this changes the bulk answer. 

Based on the idea of the purely open ensemble, a purely open tensor model was constructed in \cite{Jafferis:2025yxt}, where the Feynman diagrams are built from the open $6j$ manifold and the open ``pillow'' manifold (which can be obtained from the open $6j$ manifold). The tensor model is therefore a model of 3d triangulations. In particular, the Schwinger-Dyson equation of the tensor model is then an equation relating different ways of triangulating a given manifold. 

Let us also comment on an interesting asymmetry regarding tetrahedral decomposition in the purely closed sector versus the purely open sector. When all states are above the black hole threshold, the finite geometries for the purely open case are constructed only from the first type of tetrahedra in Figure~\ref{fig:tetraall}, whereas the building blocks for the purely closed case consist of the complete list. Conceptually, this discrepancy arises because the building blocks for the closed sector of the open-closed Virasoro TQFT are not the open $6j$ manifold, but rather the closed $6j$ manifold and the so-called knotted handcuff manifold. That manifolds contributing to closed OPE statistics can nevertheless be constructed solely from the open $6j$ manifold follows from the open–closed relation and from the fact that both EOW boundaries and OPE boundaries satisfy Neumann boundary conditions.

\section*{Acknowledgements}
It is a pleasure to thank Tom Hartman, Janet Hung, Liza Rozenberg, Zixia Wei, Cynthia Yan, and Mengyang Zhang for helpful discussions. 
DLJ acknowledges support by the Simons Investigator in Physics Award MP-SIP-0001737 and U.S. Department of Energy grant DE-SC0007870. DW acknowledges support by NSF grant PHY-2207659 and the Simons Collaboration on Celestial Holography. 

\appendix

\bibliographystyle{JHEP}
\bibliography{library}

\end{document}